\documentclass[english]{paper}
\usepackage[T1]{fontenc}
\usepackage[latin9]{inputenc}
\usepackage{float}
\usepackage{amsmath}
\usepackage{amssymb}
\usepackage{graphicx}
\usepackage{esint}

\makeatletter
\newcommand{\lyxaddress}[1]{
\par {\raggedright #1
\vspace{1.4em}
\noindent\par}
}

\makeatother

\usepackage{babel}
\begin{document}

\title{Chaos and Nonlinear Dynamics in a Quantum Artificial Economy}

\author{Carlos Pedro Gonçalves}

\institution{{\footnotesize Instituto Superior de Ciências Sociais e Políticas
(ISCSP) - Technical University of Lisbon}}

\maketitle

\lyxaddress{E-mail: cgoncalves@iscsp.utl.pt}
\begin{abstract}
Chaos and nonlinear economic dynamics are addressed for a quantum
coupled map lattice model of an artificial economy, with quantized
supply and demand equilibrium conditions. The measure theoretic properties
and the patterns that emerge in both the economic business volume
dynamics' diagrams as well as in the quantum mean field averages are
addressed and conclusions are drawn in regards to the application
of quantum chaos theory to address signatures of chaotic dynamics
in relevant discrete economic state variables.\end{abstract}

\section{Introduction}

Economic systems, worked from complex systems' science and nonlinear
dynamics \cite{key-1,key-4,key-5,key-6,key-7,key-18,key-21}, can
be approached in terms of either continuous state or discrete state
systems \cite{key-27}. Discrete state systems' modelling tools, such
as cellular automata, address the systems' state transition in terms
of discrete state maps, while continuous state systems are addressed
in terms of continuous state equations of motion, either in discrete
time (dynamical maps) or continuous time (differential equations).

The fact that, in most industries%
\footnote{By industry it is understood any economic activity whose purpose is
the creation of wealth. %
}, the transactioned quantities are discrete, might seem to render
support to discrete state rather than continuous state approaches.
The present work's proposal is that there may be a bridge between
the two approaches, such that, through the application of the mathematical
formalism of quantum mechanics to repeated (iterated) business games,
one is able to integrate continuous state nonlinear dynamical systems'
models with discrete states.

The resulting approach leads to an application of quantum chaos within
mathematical economics. By quantum chaos it is understood here the
quantum behavior of a dynamical system which on average is classically
chaotic \cite{key-2,key-23}. Since, in this case, we are dealing
with economic systems, we provide, in \textbf{section 2}., for a quantum
game theoretical framework within which quantized behavior in economic
equilibrium can be understood, including an economic-based interpretation
of the ket vectors in terms of risk pricing and a quantum game equilibrium
condition that leads to a coherent state solution. 

In \textbf{section 3.}, we integrate the approach of \textbf{section
2}. with an evolutionary game of an artificial economy, formalized
in terms of a quantum coupled map lattice with chaotic dynamics in
the quantum averages. The measure-theoretic properties of the resulting
(quantum) dynamical system as well as major dynamical features are
addressed.

In \textbf{section 4.}, we conclude with a reflection, in light of
\textbf{section 3.}'s results, on the application of quantum chaos
to nonlinear mathematical economics.

\section{Quantum Economics of Supply and Demand}

Let us consider an economy comprised of a population of $N$ competing
companies, such that, for each company, there are the following (local)
linear supply and demand functions:

\begin{equation}
Q^{s}\left(p_{t}(i)\right)=s\left(p_{t}(i)-\bar{p}\right)
\end{equation}
\begin{equation}
Q^{d}\left(p_{t}(i)\right)=d_{t}(i)-kp_{t}(i)
\end{equation}
where $i$ indexes a company and $t$ indexes a given discrete transaction
round for the economy, $\bar{p}$ is the economy-wide smallest price
at which any company is willing to sell its supply, $k$ and $s$
are fixed and characteristic of the economy, while the autonomous
demand $d_{t}(i)$ is company-specific and dynamic, such that equilibria
change by displacements of the (entire) demand line.

Each transaction round has a finite fixed duration and all transactions
are assumed to take place at the end of the round and obey the equilibrium
condition:

\begin{equation}
Q^{d}\left(p_{t}(i)\right)=Q^{s}\left(p_{t}(i)\right)=Q_{t}(i)
\end{equation}
where $Q_{t}(i)$ are the equilibrium quantities. Letting $\phi:=\frac{1}{k}+\frac{1}{s}$
and $\theta:=\frac{\bar{p}}{\phi}$, solving for the equilibrium price,
we obtain: 
\begin{equation}
Q_{t}(i)=\frac{d_{t}(i)}{\phi}+\theta
\end{equation}
Now, assuming that the quantities bought and sold are discrete undividable
units, we must have that $Q_{t}(i)$ can only assume discrete values,
thus, to address such a discrete demand we must assume a quantization
scheme such that we replace the dynamical variable $Q_{t}(i)$ by
an operator $\hat{Q}(i)$ on the economy's Hilbert space $\mathcal{H}_{Economy}$,
which is assumed to be spanned by the basis $\left\{ \left|n_{1},n_{2},...,n_{N}\right\rangle :n_{i}\in\mathbb{N}_{0},i=1,2,...,N\right\} $
and such that:
\begin{equation}
\hat{Q}(i)=\frac{\hat{d}(i)}{\phi}+\theta
\end{equation}

\begin{equation}
\hat{d}(i)=\hat{a}(i)^{\dagger}\hat{a}(i)\phi
\end{equation}

\begin{equation}
a^{\dagger}(i)\left|...,n_{i},...\right\rangle =\sqrt{n_{i}+1}\left|...,n_{i}+1,...\right\rangle 
\end{equation}
\begin{equation}
a(i)\left|...,n_{i},...\right\rangle =\sqrt{n_{i}}\left|...,n_{i}-1,...\right\rangle 
\end{equation}

\begin{equation}
\left[a(i),a^{\dagger}(j)\right]=\delta_{i,j}
\end{equation}
\begin{equation}
a_{i}\left|...,n_{i}=0,...\right\rangle =0
\end{equation}
thus, replacing Eq.(6) in Eq.(5), we obtain the operator for the equilibrium
quantities:
\begin{equation}
\hat{Q}(i)=\hat{a}(i)^{\dagger}\hat{a}(i)+\theta
\end{equation}
where $\hat{a}(i)^{\dagger}$ and $\hat{a}(i)$ are interpreted as
quantity raising and lowering operators for the company $i$, which,
from Eqs. (7) to (10), behave like bosonic creation and annihilation
operators. The spectrum of each company's quantities' operator is,
then, given by:
\begin{equation}
\hat{Q}(i)\left|...,n_{i},...\right\rangle =\left(n_{i}+\theta\right)\left|...,n_{i},...\right\rangle 
\end{equation}
where $\theta$ has to assume an integer value, since quantities are,
by assumption, expressed in whole numbers. In this way, $\theta$
is the smallest equilibrium quantity that can be sold in the market
for any given company's supply.

We may now introduce the complete projector set onto the basis of
the economy's Hilbert space $\left\{ \hat{P}_{\boldsymbol{n}}=\left|\boldsymbol{n}\right\rangle \left\langle \boldsymbol{n}\right|:\left|\boldsymbol{n}\right\rangle =\left|n_{1},n_{2},...,n_{N}\right\rangle ;n_{i}\in\mathbb{N}_{0},i=1,2,...,N\right\} $,
each such projector has the matrix form of a quantum Arrow-Debreu
security \cite{key-12}, which pays one unit of numeraire in the corresponding
equilibrium state and zero units in all the others, assuming this
analogy, we can bring to the quantum game setting an economic interpretation
that fully integrates the quantum formalism in the economic framework
to which it is being applied.

Therefore, at each round of the game, the economy is assumed to be
characterized by a normalized \emph{ket} vector $\left|\Psi(t)\right\rangle \in\mathcal{H}_{Economy}$,
such that:
\begin{equation}
\left|\Psi(t)\right\rangle =\frac{1}{\sqrt{B}}\sum_{\boldsymbol{n}}\psi\left(\boldsymbol{n},t\right)\left|\boldsymbol{n}\right\rangle 
\end{equation}
where the round $t$ amplitude $\psi(\boldsymbol{n},t)=\left\langle \boldsymbol{n}|\Psi(t)\right\rangle $
is interpreted as the beginning-of-round infimum selling price amplitude
for the end-of-round risk exposure $\hat{P}_{\boldsymbol{n}}$, or
\emph{upper prevision amplitude}, while the conjugate amplitude $\psi(\boldsymbol{n},t)^{*}=\left\langle \Psi(t)|\boldsymbol{n}\right\rangle $
is the beginning-of-round supremum buying price amplitude for the
end-of-round risk exposure $\hat{P}_{\boldsymbol{n}}$, or \emph{lower
prevision amplitude}, in this way, the price for risk is $\left\langle \boldsymbol{n}\left|\hat{P}_{\boldsymbol{n}}\right|\boldsymbol{n}\right\rangle =|\psi(\boldsymbol{n},t)|^{2}=\psi(\boldsymbol{n},t)^{*}\psi(\boldsymbol{n},t)$,
this is a quantum Arrow-Debreu price or quantum state-contingent price
\cite{key-12}. From the normalization condition, we have $\sum_{\boldsymbol{n}}|\psi(\boldsymbol{n},t)|^{2}=B$,
where $B$ is a discount factor for the duration of a game round.

In this way, $\frac{|\psi(\boldsymbol{n},t)|^{2}}{B}$ are the end
of round Arrow-Debreu capitalized prices and, from the way in which
they are calculated, can be stated to satisfy the following relation:
\begin{equation}
\underline{P}\left[\hat{P}_{\boldsymbol{n}}\right]=\frac{|\psi(\boldsymbol{n},t)|^{2}}{B}=\overline{P}\left[\hat{P}_{\boldsymbol{n}}\right]
\end{equation}
where $\underline{P}\left[\hat{P}_{\boldsymbol{n}}\right]$ and $\overline{P}\left[\hat{P}_{\boldsymbol{n}}\right]$
are, respectively, the \emph{lower} and \emph{upper previsions} on
the alternative $\hat{P}_{\boldsymbol{n}}$ \cite{key-25}, therefore,
each $\hat{P}_{\boldsymbol{n}}$ can be stated to be priceable such
that the \emph{infimum acceptable selling price} (\emph{upper prevision})
and the \emph{supremum acceptable buying price} (\emph{lower prevision})
for a transaction on a risk exposure to $\hat{P}_{\boldsymbol{n}}$
both coincide with $\frac{|\psi(\boldsymbol{n},t)|^{2}}{B}$. Thus,
if a bet were made on the risk exposure $\hat{P}_{\boldsymbol{n}}$,
then, its end-of-round fair price would be $\frac{|\psi(\boldsymbol{n},t)|^{2}}{B}$
synthesizing, simultaneously, a price and a likelihood.

While classical probabilities are pure numbers, a probability with
classical additivity that comes from the coincidence between an \emph{upper}
and \emph{lower prevision} can be expressed in monetary units, such
that it expresses, in the pricing, a statement about the risk and,
thus, the likelihood of an event, the fair price $\frac{|\psi(\boldsymbol{n},t)|^{2}}{B}$
is, therefore, an expression on an exposure to a risk situation which
reflects a likelihood systemic evaluation, that is, the fair pricing
of risk must reflect the the risky event's likelihood, and, therefore,
we are dealing with a risk cognition as a systemic cognitive synthesis
about risk, enacted by a complex adaptive system (in the present case,
it is the system of companies).

This pricing is a systemic cognitive synthesis, as per imprecise probabilities
theory, and not necessarily means that we are assuming the actual
trading of Arrow-Debreu securities, it does mean that agents, in this
case, companies, cognitively evaluate the alternatives and are able
to assign to them a fair value for risk exposure \cite{key-12}, which
can be recognized by any agent with the same data regarding the risk
systemic situation, in this case, the risk is the business economic
risk associated with different alternative equilibrium quantities
and, therefore, different business volumes.

Each company's pricing of the game can be obtained from $\left|\Psi(t)\right\rangle $
by introducing the operator chains:

\begin{equation}
\hat{C}_{n_{i}}=\sum_{...n_{i-1},n_{i+1}...}\left|...,n_{i-1},n_{i},n_{i+1},...\right\rangle \left\langle ...,n_{i-1},n_{i},n_{i+1},...\right|
\end{equation}
with $\sum_{n_{i}}\hat{C}_{n_{i}}=\hat{1}$. For each alternative
equilibrium state configuration of company $i$, the chain is obtained
from a sum over all of the other alternatives for the rest of the
companies, such chains are called coarse-grained chains in the decoherent
histories approach to quantum mechanics \cite{key-9,key-13}, following
this approach, one may introduce the pricing functional from the expression
of the decoherence functional:
\begin{equation}
\mathcal{D}\left(m_{i},n_{i}:\left|\Psi(t)\right\rangle \right)=\left\langle \Psi(t)\left|\hat{C}_{m_{i}}^{\dagger}C_{n_{i}}\right|\Psi(t)\right\rangle 
\end{equation}
we, then, have, for each company:
\begin{equation}
\mathcal{D}\left(m_{i},n_{i}:\left|\Psi(t)\right\rangle \right)=0,\:\forall m_{i}\neq n_{i}
\end{equation}
from condition (17), it follows that the strategy matrix, defined
analogously to the decoherence matrix from the decoherent histories
approach, is a diagonal matrix and can be expanded as a linear combination
of the projectors for each company as follows:
\begin{equation}
\mathbf{D}_{i}\left(\left|\Psi(t)\right\rangle \right)=\sum_{n_{i}}\mathcal{D}\left(n_{i},n_{i}:\left|\Psi(t)\right\rangle \right)\hat{P}_{n_{i}}
\end{equation}
which has the mathematical expression of a game mixed strategy \cite{key-20}.
In particular, $\mathbf{D}_{i}\left(\left|\Psi(t)\right\rangle \right)$
can be regarded as points in a simplex whose vertices are the projectors
$\hat{P}_{n_{i}}$. Likewise, the weights $\mathcal{D}\left(n_{i},n_{i}:\left|\Psi(t)\right\rangle \right)$
satisfy a condition of equality with respect to \emph{upper} and \emph{lower
previsions}:
\begin{equation}
\underline{P}\left[\hat{P}_{n_{i}}\right]=\mathcal{D}\left(n_{i},n_{i}:\left|\Psi(t)\right\rangle \right)=\overline{P}\left[\hat{P}_{n_{i}}\right]
\end{equation}
which follows from condition (17) and fact that \emph{lower previsions}
are superadditive (null or constructive interference) and \emph{upper
previsions} are subadditive (null or destructive interference) only
coinciding with each other (null interference, or, in the present
case, pricing additivity) when the decoherence condition of equation
(17) holds. Decoherence, in this case, implies priceability of risk
exposure as well as the ability to assign a valuation scheme for the
company's game position present value at each transaction round, indeed,
if we assume that, at each round, a company needs to invest an amount
$I_{i}$ to support its economic activity, then, since the $\mathcal{D}\left(n_{i},n_{i}:\left|\Psi(t)\right\rangle \right)$
are end-of-round prices, the company's game position present value
at the beginning of a round $t$ can be determined as \cite{key-3,key-12}:
\begin{equation}
-I_{i}+B\left\langle \hat{Q}(i)\right\rangle _{\Psi(t)}=-I_{i}+B\sum_{n_{i}}\left(n_{i}+\theta\right)\mathcal{D}\left(n_{i},n_{i}:\left|\Psi(t)\right\rangle \right)
\end{equation}
Assuming that companies manage business volatility, a game equilibrium
solution$\left|\Psi(t)\right\rangle $, for each round, can be obtained
through an appropriate game payoff function, in this case, considering
$A_{i}=a_{i}-\left\langle a_{i}\right\rangle _{\Psi(t)}$ and $A_{i}^{\dagger}=a_{i}^{\dagger}-\left\langle a_{i}^{\dagger}\right\rangle _{\Psi(t)}$
, we may introduce the following quantum volatility risk measure:
\begin{equation}
R(i;\left|\Psi(t)\right\rangle )=\left\langle \frac{A_{i}^{\dagger}A_{i}+A_{i}A_{i}^{\dagger}}{2}\right\rangle _{\Psi(t)}
\end{equation}
such that if $Q_{t}^{e}(i)$ is a mean evolutionarily sustainable
economic equilibrium quantity for the company at the end of round
$t$, then, we can introduce the following optimization problem:
\begin{equation}
\begin{array}{cc}
\left|\Psi(t)\right\rangle = & \min_{all\left|\Phi\right\rangle }R(i;\left|\Phi\right\rangle )\\
s.t. & \left\langle a\right\rangle _{\Psi(t)}=\sqrt{Q_{t}^{e}(i)-\theta}
\end{array}
\end{equation}
The minimization is taken over all of the normalized \emph{kets} in
the economy's Hilbert space, and it leads to a coherent state solution
for each company \cite{key-10}:
\begin{equation}
a_{i}\left|\Psi(t)\right\rangle =\alpha_{t}(i)\left|\Psi(t)\right\rangle 
\end{equation}
on the other hand, the restriction leads to the following result:
\begin{equation}
\alpha_{t}(i)=\sqrt{Q_{t}^{e}(i)-\theta}
\end{equation}

The round $t$ quantum game equilibrium solution is, thus, given by
the tensor product of $N$ coherent states:
\begin{equation}
\left|\Psi(t)\right\rangle =\bigotimes_{i=1}^{N}\left|\alpha_{t}(i)\right\rangle 
\end{equation}
\begin{equation}
\left|\alpha_{t}(i)\right\rangle =\exp\left(-\frac{\alpha_{t}(i)^{2}}{2}\right)\sum_{n=0}^{+\infty}\frac{\alpha_{t}(i)^{n}}{\sqrt{n!}}\left|n\right\rangle 
\end{equation}
which leads to the resulting mixed strategies:
\begin{equation}
\mathbf{D}_{i}\left(\left|\Psi(t)\right\rangle \right)=\sum_{n_{i}=0}^{\infty}\frac{\exp\left(-\frac{Q_{t}^{e}(i)-\theta}{2}\right)\left(Q_{t}^{e}(i)-\theta\right)^{n_{i}}}{n_{i}!}\hat{P}_{n_{i}}
\end{equation}
There is a unitary transition from round to round, for each company,
that links two consecutive quantum game equilibrium solutions through:
\begin{equation}
\hat{U}_{i}\left(Q_{t}^{e}(i),t,t-1\right)=\exp\left[\sqrt{Q_{t}^{e}(i)-\theta}\left(a_{i}^{\dagger}-a_{i}\right)\right]\exp\left[-\sqrt{Q_{t-1}^{e}(i)-\theta}\left(a_{i}^{\dagger}-a_{i}\right)\right]
\end{equation}
At the end of each transaction round, for each company, the equilibrium
quantities $Q_{t}(i)$ follow a random Poisson distribution such that,
from equation (27) and the coincidence between the \emph{upper} and
\emph{lower previsions} (decoherence), we have the probabilities:
\begin{equation}
P\left[Q_{t}(i)=n_{i}+\theta\right]=\left\langle n_{i}\right|\mathbf{D}_{i}\left(\left|\Psi(t)\right\rangle \right)\left|n_{i}\right\rangle 
\end{equation}
Quantum chaos can take place, within such a game, whenever $Q_{t}^{e}(i)$
follows a chaotic dynamics. It is to this point that we now turn.

\section{Quantum Chaos in an Artificial Economy}

Assuming the previous section's formalism, let us now consider that
each company is characterized by core business dimensions which include
mission statement and business concept that define the company's core
business strategic profile and are specific to each company, such
that a company's ID-code is introduced in the form of a binary string
$\sigma_{i}$ of length $k$, there being $N=2^{k}$ companies in
the economy, where $\sigma_{i}$ can be considered as a company's
address in the core business dimensions' hypercubic lattice%
\footnote{The lattice vertices correspond to the $\sigma_{i}$ and each lattice
connection links two vertices that differ by just one digit.%
}. Then, the following dynamics is introduced for $Q_{t}^{e}(i)$:
\begin{equation}
Q_{t}^{e}(i):=Q_{e}\left(x_{t}(\sigma_{i})\right)=\bar{Q}+\nu x_{t}(\sigma_{i})
\end{equation}
\begin{equation}
x_{t}(\sigma_{i})=\Phi\left(x_{t-1}(\sigma_{j})\right)=\left(1-\varepsilon_{1}-\varepsilon_{2}\right)f_{b}\left(x_{t-1}(\sigma_{i})\right)+\frac{\varepsilon_{1}}{K}\sum_{j=1}^{K}f\left(x_{t-1}\left(\sigma_{j}(i)\right)\right)+\frac{\varepsilon_{2}}{N}\sum_{k=1}^{N}f\left(x_{t-1}\left(\sigma_{k}\right)\right)
\end{equation}
\begin{equation}
f_{b}\left(x_{t-1}(\sigma_{i})\right)=(1-\delta)\left[1-bx_{t-1}(\sigma_{i})^{2}\right]+\delta M_{t-1}(i),\, M_{t}(i)=\frac{Q_{i}(t)}{\sum_{j=1}^{N}Q_{j}(t)}
\end{equation}
In equation (30), $Q_{t}^{e}(i)$ is a function of a fixed industry-wide
quantity $\bar{Q}$ and a variable company-specific term $x_{t}(\sigma_{i})$,
rescaled by $\nu$, the variable $x_{t}(\sigma_{i})$ represents the
company's core business fitness, defined in terms of its ability to
satisfy the needs of the consumer market. In this case, we are dealing
with a fitness field on the core business dimensions' hypercubic lattice.

In equation (30), each company's core business fitness $x_{t}(\sigma_{i})$
depends upon a coupled nonlinear map $\Phi\left(x_{t-1}(\sigma_{j})\right)$
comprised of two couplings, the first is a coupling to all of the
$K$ companies in the lattice that differ by just one digit from the
\emph{i}-th company in their address at the core business dimensions'
hypercubic lattice, thus, $\sigma_{j}(i)$ corresponds to the company
that differs from $i$ by its $j$-th digit. This local coupling represents
demand flow from other companies that are similar in profile in terms
of their core business, therefore, we are dealing with a consumer
diffusion-like adaptive walk in the hypercubic lattice, such that
the strength of the coupling $\varepsilon_{1}$ is similar to a single-point
mutation in evolutionary biology, if we let $\varepsilon_{2}=0$,
the model for the dynamics of $x_{t}(\sigma_{i})$ is, indeed, akin
to Kaneko's \emph{self-organizing genetic algorithms} \cite{key-15,key-16},
where $\varepsilon_{1}$ plays the role of a mutation rate, in the
present business game case, the process corresponds to consumers flipping
their consumption pattern.

The second coupling $\varepsilon_{2}$ corresponds to a global industry-wide
competition term, if we let $\varepsilon_{1}=0$, then, we obtain
a globally coupled map. For $\varepsilon_{1}=0$ and $\varepsilon_{2}\neq0$,
we have an economy with supply differentiation interplaying with adaptive
behavior on the part of the demand, as well as global synchronization
components associated with global market-wide competition.

The nonlinear map $f_{b}$, following equation (32), is a quadratic
map%
\footnote{The one-humped family examples (the quadratic map and the logistic
map) appear recurrently in models of economic growth \cite{key-18},
therefore, the choice of the quadratic leaves room for adaptation
of the model to other approaches, which is effective, for comparison
sake.%
}, coupled to the previous round's company's market share that, in
turn, is calculated from the previous round's economic equilibrium
quantities, which are a (quantum) probabilistic result, as per previous
section's formalism.

When $\varepsilon_{1}=\varepsilon_{2}=\delta=0$, and $b$ leads to
an attracting fixed point dynamics, each company is characterized
by the same coherent state solution, we have a case of independent
and identically distributed noise both in space as well as in time.
When $b$ is in an attracting cycle region, there is a periodic dynamics
for the companies' kets, such that each company's business dynamics
is no longer described by identically distributed noise.

When $b$ is in the chaotic region, then, it becomes possible to provide
for a quantum statistical description of the game solutions in their
relation with the classical Perron-Frobenius operator statistical
description of chaos. Thus, at each round of the game, one may introduce
a density operator for each company, given by:

\begin{equation}
\hat{\rho}\left[x_{t}(\sigma_{i})\right]=\left|\sqrt{Q_{e}\left(x_{t}(\sigma_{i})\right)-\theta}\right\rangle \left\langle \sqrt{Q_{e}\left(x_{t}(\sigma_{i})\right)-\theta}\right|
\end{equation}
which is a pure state density with quantum evolution rule:

\begin{equation}
\hat{\rho}\left[x_{t}(\sigma_{i})\right]=\hat{U}_{i}\left(Q_{e}\left(x_{t}(\sigma_{i})\right),t,t-1\right)\hat{\rho}\left[x_{t-1}\left(\sigma_{i}\right)\right]\hat{U}_{i}\left(Q_{e}\left(x_{t}(\sigma_{i})\right),t,t-1\right)^{\dagger}
\end{equation}
One can address the sequence of density operators as an orbit in the
space of quantum solutions for the economic game, such that, denoting
the Perron-Frobenius operator by $\hat{E}_{PF}$, we have: 
\begin{equation}
\begin{aligned}\hat{\rho}\left[x_{t}(\sigma_{i})\right]=\\
 & =\int dx\hat{\rho}\left(x\right)\hat{E}_{PF}\delta\left(x-x_{t-1}(\sigma_{i})\right)=\\
 & =\int dx\hat{\rho}\left(x\right)\delta\left(x-x_{t}(\sigma_{i})\right)
\end{aligned}
\end{equation}
where $\delta\left(x-x_{t-1}(\sigma_{i})\right)$ is Dirac's delta
function. Replacing (35) in (34), we obtain a relation between the
classical state transition of the nonlinear dynamical system and the
quantum game: 
\begin{equation}
\begin{aligned}\hat{U}_{i}\left(Q_{e}\left(x_{t}(\sigma_{i})\right),t,t-1\right)\hat{\rho}\left[x_{t-1}\left(\sigma_{i}\right)\right]\hat{U}_{i}\left(Q_{e}\left(x_{t}(\sigma_{i})\right),t,t-1\right)^{\dagger}=\\
=\int dx\hat{\rho}\left(x\right)\hat{E}_{PF}\delta\left(x-x_{t-1}(\sigma_{i})\right)
\end{aligned}
\end{equation}
In the case of ergodic chaos, there is an invariant density which
is a fixed point of the Perron-Frobenius operator, that is, $\rho_{PF}\left(x\right)=\hat{E}_{PF}\rho_{PF}\left(x\right)$,
which leads to the coherent state ensemble statistical picture result:
\begin{equation}
\begin{aligned}\int dx\rho_{PF}\left(x\right)\left|\sqrt{Q_{e}\left(x\right)-\theta}\right\rangle \left\langle \sqrt{Q_{e}\left(x\right)-\theta}\right|=\\
=\int dx\hat{E}_{PF}\rho_{PF}\left(x\right)\left|\sqrt{Q_{e}\left(x\right)-\theta}\right\rangle \left\langle \sqrt{Q_{e}\left(x\right)-\theta}\right|
\end{aligned}
\end{equation}
with $Q_{e}(x):=\bar{Q}+\nu x$. In the present case, for $b=2$,
for $\varepsilon_{1}=\varepsilon_{2}=\delta=0$, the orbits for $x_{t}(\sigma_{i})$
are ergodic, such that, while, the sequence of quantum game solutions
for each company are described by a sequence of pure state density
operators, there is the statistical stability of the invariant density
description that obeys (37) and is given by:
\begin{equation}
\int dx\frac{1}{\pi\sqrt{\left(4-x^{2}\right)}}\left|\sqrt{Q_{e}\left(x\right)-\theta}\right\rangle \left\langle \sqrt{Q_{e}\left(x\right)-\theta}\right|
\end{equation}
this is the same quantum statistical state in the thermodynamic limit
of $N\rightarrow+\infty$, under the invariant density as statistical
distribution. Thus, the classical ergodic chaotic dynamics of the
quadratic map leads to a statistical description of the quantum economic
dynamics in terms of an ensemble of coherent states with the invariant
density playing the role of the statistical coherent state density.

One must be careful however, in the statistical interpretation, since
the ensemble state (38) cannot be interpreted as a probability distribution
for different coherent states, since the coherent states are not orthogonal,
the final result of (38) is an ensemble of coherent states quantum
game solutions \cite{key-11,key-24}.

Once spatial coupling is assumed, however, the fixed point solution
of the Perron-Frobenius operator becomes unstable such that the quantum
stability of (38) changes, this already takes place when $\varepsilon_{2}\neq0$
and $\delta=\varepsilon_{1}=0$.

The global coupling case of $\varepsilon_{2}\neq0$ and $\delta=\varepsilon_{1}=0$
is conceptually linked to a perfect competition model such that the
higher the value of $\varepsilon_{2}$ is, the lower is the differentiation
between the companies' supply, the instability, in this case, is linked
to the collective behavior of the companies' core business fitness
mean field $\frac{1}{N}\sum_{k=1}^{N}x_{t}\left(\sigma_{k}\right)$,
the mean field dynamics and the density for the global coupling case
are mutually related, such that, with a change in the mean field,
the Perron-Frobenius fixed point is changed and a change in the Perron-Frobenius
invariant density, in turn, leads to a change in the mean field, these
changes take place with some delay, leading to oscillatory dynamics
in the ensemble over the coherent states solution, a result which
comes out of the classical models \cite{key-16}.

For the full model dynamics with $\varepsilon_{1}\neq0$, $\varepsilon_{2}\neq0$
and $\delta\neq0$, the perfect competition model gives way to an
evolutionary race between companies for market share, with: (1) local
competition dynamics between companies that are close to each other
in their core business strategic dimensions ($\varepsilon_{1}\neq0$);
(2) market share feedback effects upon a company's core business fitness
($\delta\neq0$) and (3) global competitiveness industry-wide race
($\varepsilon_{2}\neq0$). All of these elements make the artificial
economy much closer to the strategically differentiated industries,
with networked coevolving economic dynamics, that characterize current
market economies.

When both supply differentiation and global coupling are considered,
complex behavior emerges in the chaotic regime. In Fig.1, the sequence
of quantities sold for a Netlogo simulation of the economy with $N=256$
companies is plotted in a cellular automaton-like business volume
dynamics diagram with each line corresponding to a transaction round
and each column to a company, grey-scale rendering was used for the
quantities sold, using Fraclab, for $b=2$ and low coupling $\varepsilon_{1}=\varepsilon_{2}=0.01$
(Fig.1, top) as well as for high local coupling $\varepsilon_{1}=0.3$
with low global coupling $\varepsilon=0.01$ (Fig.1, bottom).

For low local and global coupling, each company is highly differentiated
from its competitors, such that turbulent chaotic dynamics dominates
the core business fitness dynamics, which can be seen in the top diagram
of Fig.1. On the other hand, when the local coupling is increased,
such that there is high competitiveness between companies with similar
core business strategic dimensions, a periodic business cycle pattern
emerges from among the turbulence, thus, periodicity in the business
dynamics emerges within a chaotic noisy background.

To address the emerging patterns, in both diagrams, it is useful to
apply a lacunarity analysis. The lacunarity measure for a two-dimensional
image describes the dispersion of the luminosities present in the
windows of size $\delta$, denoted by $L_{\delta}$, with respect
to the mean luminosity in windows of size $\delta$, denoted by $M_{\delta}$,
as $Lac(\delta)=\left\langle \left(\frac{L_{\delta}}{M_{\delta}}-1\right)^{2}\right\rangle $,
where the brackets represent the mean over all windows of size $\delta$
\cite{key-8}. Evaluated, for different window sizes, one obtains
a lacunarity distribution.

In the present case, low luminosity corresponds to lower quantities
sold, with the reverse holding for high luminosity. Changes in luminosities,
breaks, clusterings of holes (or black spots), all of these correspond
to changes in quantities sold and, in the case of clusterings of holes,
to clustering of lower quantities sold, such that the luminosity distribution
becomes a useful business volume risk analysis tool.

In Fig.2, it is presented Fraclab-estimated lacunarity distributions
for both diagrams of Fig.1, shown in a log-log scale. There is, in
both cases, power-law scaling of lacunarities, which is a revealing
geometric signature of fractal-like structures, however, while the
power law scaling shows a straight line pattern for the first case
($\varepsilon_{1}=0.01$), for the second case ($\varepsilon_{1}=0.3$),
it shows a jagged-like pattern which is characteristic of the periodicity,
and implies that there are, simultaneously, periodic patterns and
fractal-like self-similarities in the geometric structure of the bottom
diagram of Fig.1.

The quantities sold are the discrete picture that an economist might
have of the market, such that an economist 'looking' at such a system
would approach it from the observed effects, that is, from the transactioned
quantities (economic output). The origin of the observed effects can,
in this case, be traced back to its source in the fluctuations that
take place in the quantum averages. Indeed, introducing the mean field
economic output operator for the economy:
\begin{equation}
\hat{Q}_{h}=\frac{1}{N}\left(\sum_{i=1}^{N}a_{i}^{\dagger}a_{i}+N\theta\right)
\end{equation}
we obtain the quantum average of $\hat{Q}_{h}$ at each transaction
round:
\begin{equation}
\left\langle \hat{Q}_{h}\right\rangle _{\Psi_{t}}=\frac{1}{N}\sum_{i=1}^{N}Q_{e}\left(x_{t}(\sigma_{i})\right)
\end{equation}
which assumes continous values. In Fig.3, the orbit for the quantum
average is shown for the two parameter sets of Fig.1. In the case
of $\varepsilon_{1}=0.01$, we can see how the business volume chaos
is present in the quantum average's dynamics (Fig.3, left graph).
The estimated correlation dimension is 15.05, with a time lag of 9
(minimum of the mutual information), thus, although there seems to
be a high-dimensional chaotic dynamics, in the quantum average of
the mean output, the attractor has a smaller number of dimensions
with respect to the system's size $N=256$, which means that some
synchronization is present with self-organization in a chaotic regime.

For the case of $\varepsilon_{1}=0.3$ (Fig.3, right graph), there
is still chaotic dynamics in the quantum average, but the system is
near a periodic window due to the local synchronization, this is confirmed
in the power spectrum (Fig.4), which shows evidence of noise as well
as of two frequencies that stand out, being related to the skeleton
of the periodic window.

Taking into account the overall results of the present model, it shows
how one can, with a quantum artificial economy, address different
business economic profiles of an industry and research the emerging
patterns in both the discrete economic output values as well as in
the quantum averages, linking complex patterns and dynamical signatures
at the level of the effects with their source in the quantum game.

\section{Conclusions on Quantum Chaos and Economics}

In the present work, a model of an artificial economy was built with
discrete (quantized) economic equilibrium conditions and chaotic dynamics
in the quantum averages.

The framework for an economic interpretation of the quantum formalism,
developed in \textbf{section 2.}, allowed us to develop the model
of an artificial economy, building a bridge between the discrete state
and the continuous state approaches to modelling complex economic
dynamics \cite{key-6,key-7}. 

In \textbf{section 3.}, it was shown that the quantum evolutionary
dynamics, addressed in terms of an appropriate unitary operator, and
reflecting a game equilibrium condition, allowed for the traces of
chaos in a coupled map's dynamics to be seen in the quantum probabilistic
dynamics, a result that puts into perspective the findings of noisy
chaotic signatures in economic time series of discrete dynamical variables.

Thus, the present work allows one to conjecture that noisy economic
chaos evidence, in particular in economic time series of discrete
state variables, may be effectively addressed in terms of quantum
chaos tied into quantum game theory, providing for an approach to
deal with nonlinear economic dynamics' theory in conjunction with
the conceptual framework of complex adaptive systems' science, indeed,
quantum chaos becomes a third approach to be added to the nonlinear
deterministic and nonlinear deterministic plus noise models of economic
chaos \cite{key-4,key-5}, since, by addressing evolutionary quantum
strategies we are not dealing with a \emph{plus noise approach} but,
instead, with an evolutionary systemic dynamics where probability
distributions and chaotic dynamics are interconnected with risk cognition
and business adaptive processes, thus, deepening the conceptual grounding
on complex adaptive systems' science.

\section*{Figures}

\begin{figure}[H]
\includegraphics[scale=0.8]{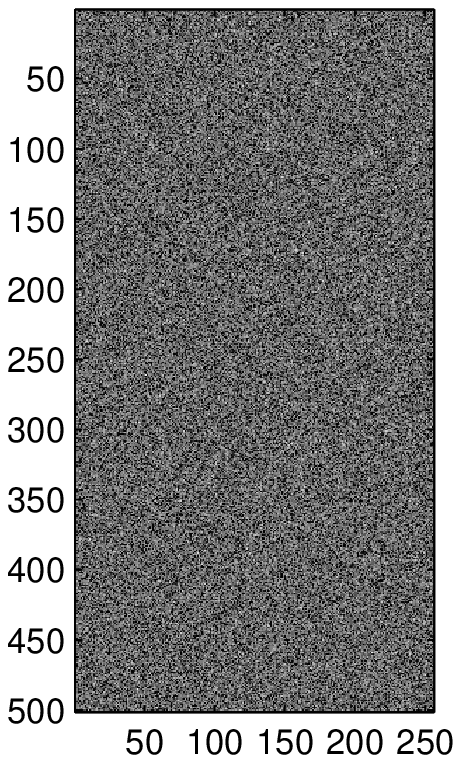}

\includegraphics[scale=0.8]{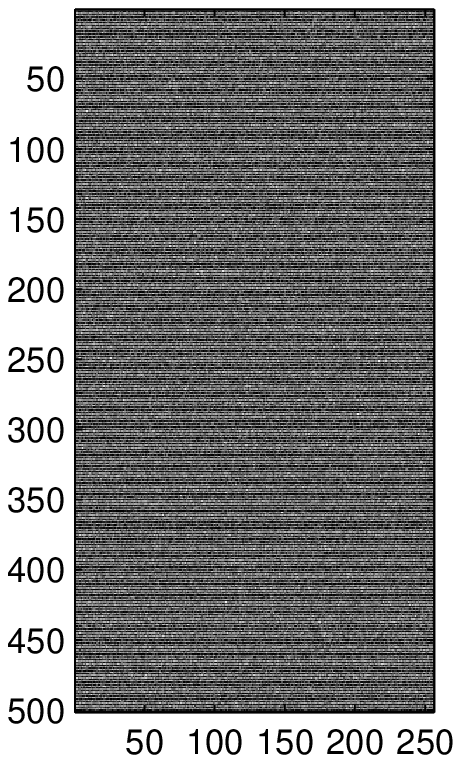}

\caption{{\small Business volume dynamics' diagrams for the quantities sold,
with $k=8$ business dimensions (i.e., $N=256$), with parameters:
$b=2$, $\theta=14$, $\bar{Q}=30$, $\nu=30$, $\varepsilon_{2}=0.04$,
$\delta=0.1$ and $\varepsilon_{1}=0.01$ (top diagram), $\varepsilon_{1}=0.3$
(bottom diagram). The grey scale is mapped automatically by the Software
Fraclab from a Netlogo simulation, the darker regions corresponding
to lower quantities sold, and lighter regions to higher quantities
sold, transaction rounds are represented in the vertical while the
spacial configuration of companies is represented horizontally. The
10,000 to 10,500 simulation steps are being shown.}}
\end{figure}

\begin{figure}[H]
\begin{centering}
\includegraphics[scale=0.4]{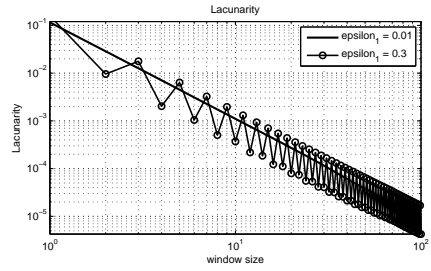}
\par\end{centering}

\caption{{\small Lacunarity distributions for the previous figure diagrams.}}
\end{figure}

\begin{figure}[H]
\begin{centering}
\includegraphics[scale=0.3]{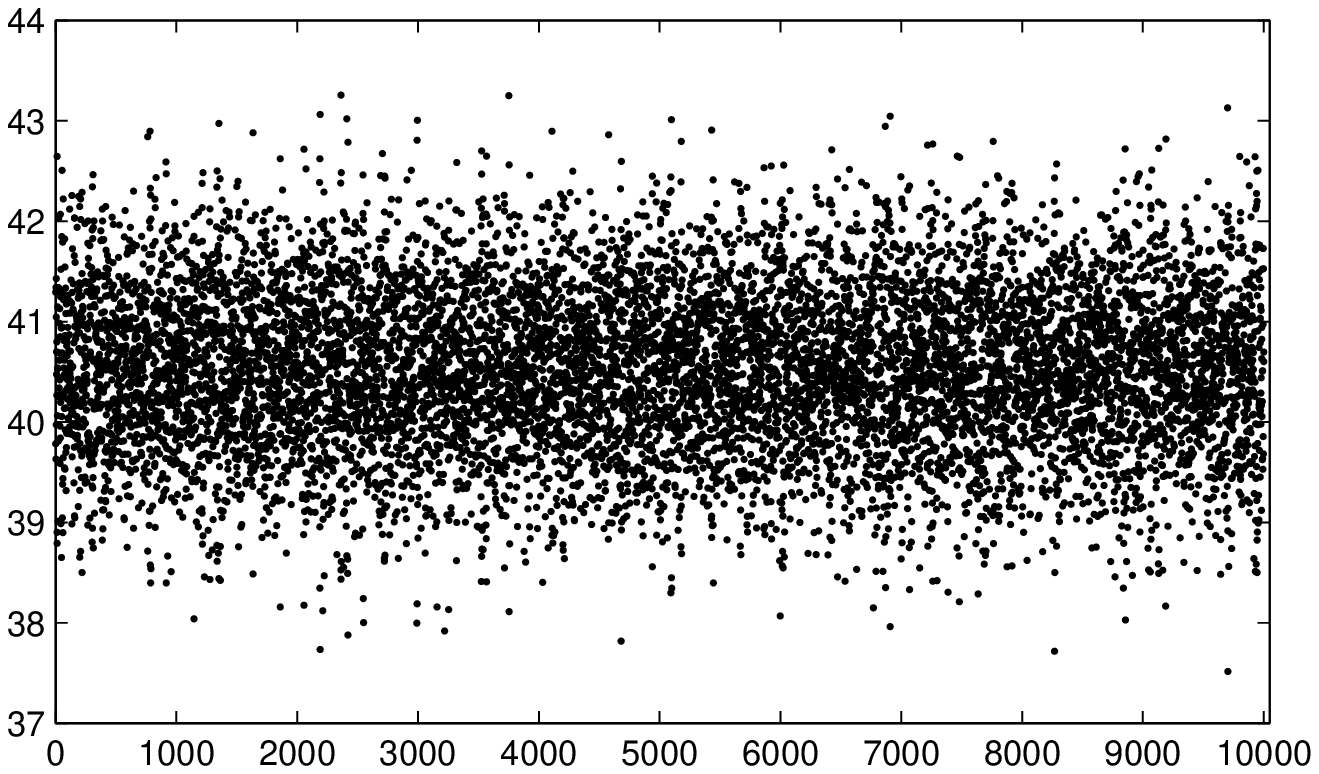}\includegraphics[scale=0.3]{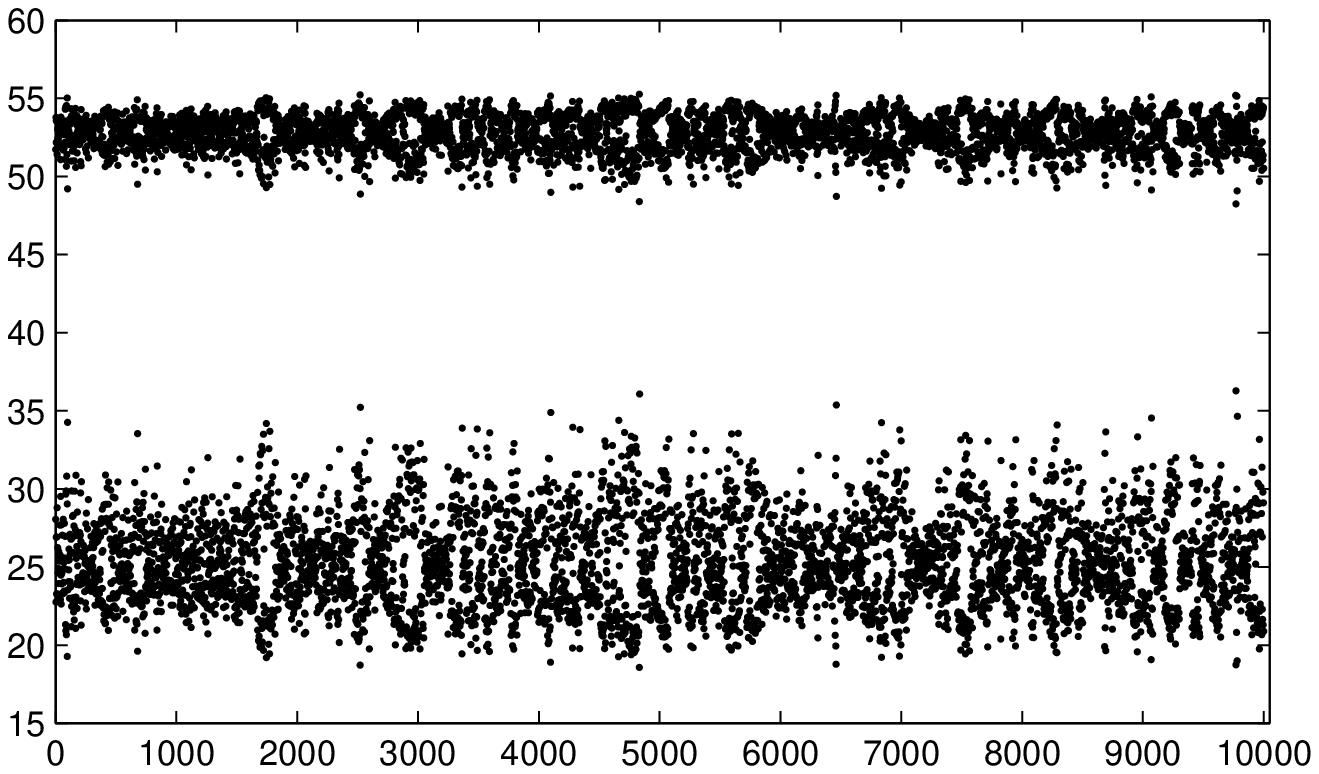}
\par\end{centering}

\caption{{\small Time series diagrams for the quantum averages $\left\langle \hat{Q}_{h}\right\rangle _{\Psi_{t}}$,
with $k=8$ business dimensions (i.e., $N=256$), with parameters:
$b=2$, $\theta=14$, $\bar{Q}=30$, $\nu=30$, $\varepsilon_{2}=0.04$,
$\delta=0.1$ and $\varepsilon_{1}=0.01$ (left diagram), $\varepsilon_{1}=0.3$
(right diagram). The 10,000 to 20,000 Netlogo simulation steps are
being shown.}}
\end{figure}

\begin{figure}[H]
\begin{centering}
\includegraphics[scale=0.3]{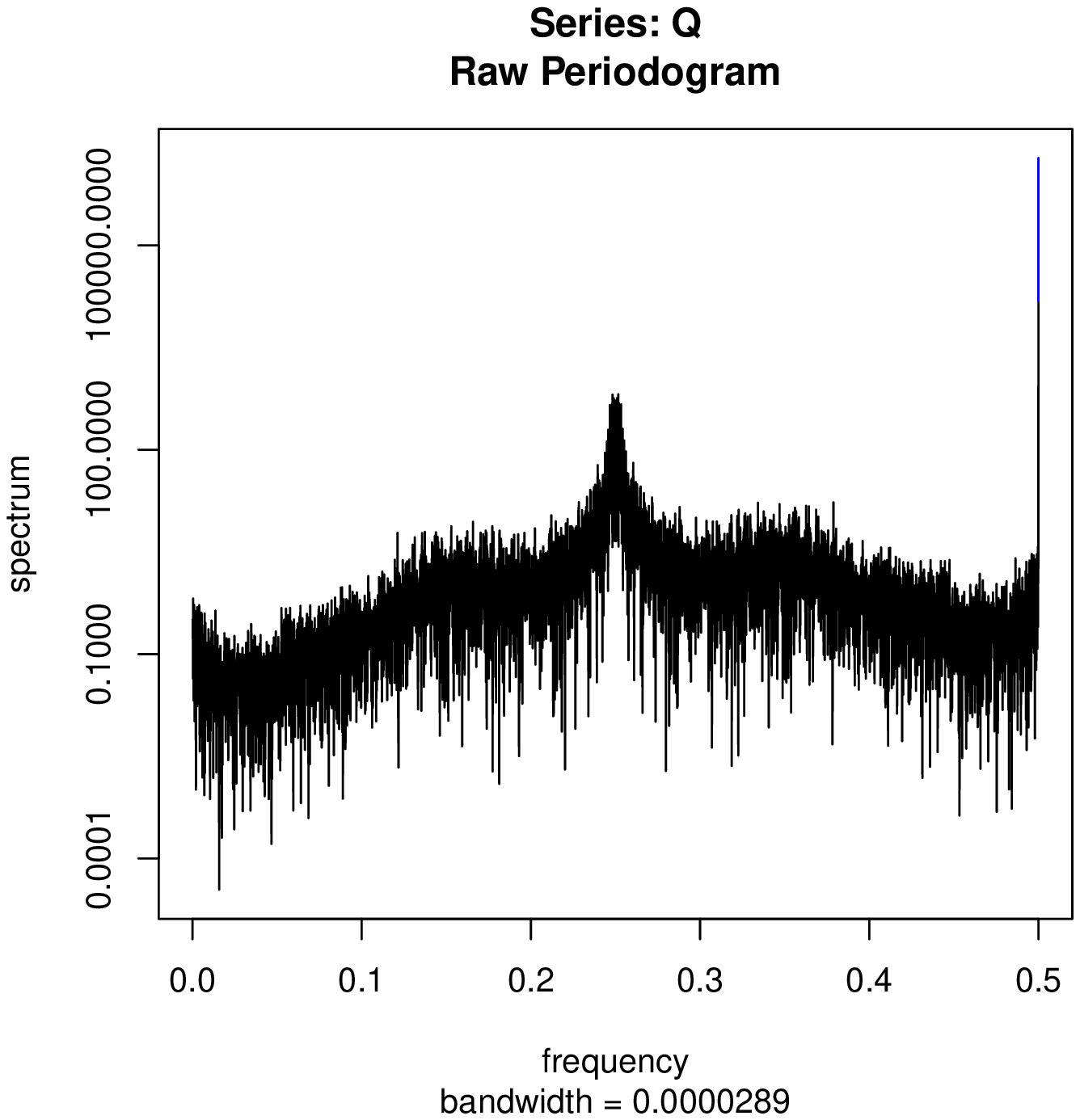}\includegraphics[scale=0.3]{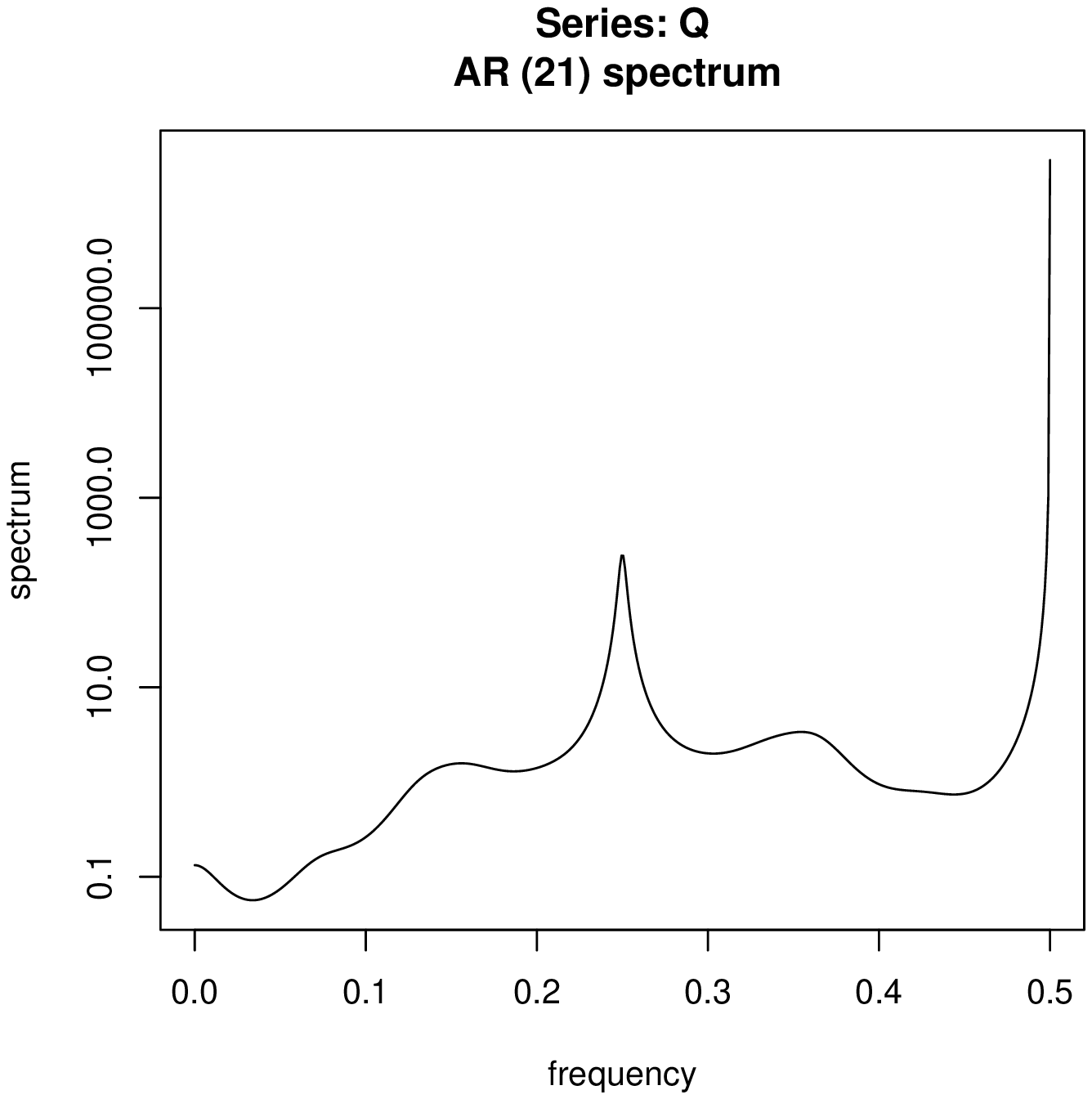}
\par\end{centering}

\caption{Raw (left) and smoothed (right) spectrograms for the previous figure
data corresponding to the simulation with $\varepsilon_{1}=0.3$.}
\end{figure}

\end{document}